\documentclass[11]{article}
\usepackage{fleqn}
\usepackage{graphicx}
\usepackage{tikz}
\usepackage{amsmath}
\usepackage{amssymb}
\usepackage{multicol}

\newtheorem{definition}{Definition}
\begin{document}
\Large
\begin{center}
{\bf Contextuality with a Small Number of Observables}
\end{center}
\vspace*{.5cm}
\large
\begin{center}
Fr\'ed\'eric Holweck$^{1}$ and Metod Saniga$^{2}$
\end{center}
\vspace*{-.3cm} 
\begin{center}

$^{1}$ IRTES/UTBM, Universit\'e de Bourgogne-Franche-Comt\'e,\\ 
F-90010 Belfort, France\\ (frederic.holweck@utbm.fr)
\vspace*{0.4cm}

$^{2}$Astronomical Institute, Slovak Academy of Sciences,\\
SK-05960 Tatransk\' a Lomnica, Slovak Republic\\
(msaniga@astro.sk)

\end{center}

\vspace*{.2cm} \noindent \hrulefill

\vspace*{.2cm} \noindent {\bf Abstract}

\noindent 
We investigate small geometric configurations that furnish observable-based proofs of the Kochen-Specker theorem. Assuming that each context consists of the same number of observables and each observable is shared by two contexts, it is proved that the most economical proofs are the famous Mermin-Peres square and the Mermin pentagram featuring, respectively, $9$ and $10$ observables, there being no proofs using less than $9$ observables. We also propose a new proof with $14$ observables forming a `magic' heptagram.
On the other hand, some other prominent small-size finite geometries, like the Pasch configuration and the prism, are shown not to be contextual.  

\vspace*{-.0cm} \noindent \hrulefill

\section{Introduction}
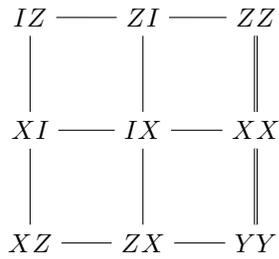
\begin{figure}[!h]
\begin{center}
\begin{tikzpicture}[scale=1.5]
\node (A) at (0,2) {$IZ$};
\node (B) at (1,2) {$ZI$};
\node (C) at (2,2) {$ZZ$};
\node (D) at (0,1) {$XI$};
\node (E) at (1,1) {$IX$};
\node (F) at (2,1) {$XX$};
\node (G) at (0,0) {$XZ$};
\node (H) at (1,0) {$ZX$};
\node (I) at (2,0) {$YY$};
\draw[-]
(A) edge (B)
(A) edge (D)
(B) edge (C)
(B) edge (E)
(C) edge [double] (F)
(D) edge (E)
(D) edge (G)
(E) edge (F)
(E) edge (H)
(F) edge [double] (I)
(G) edge (H)
(H) edge (I);
\end{tikzpicture}
\end{center}
 \caption{An illustration of the Mermin-Peres square.}\label{grid}
\end{figure}

\begin{figure}[!h]
 \begin{center}
  \begin{tikzpicture}[scale=1.3]
   \node (A) at (0,0) {$IYI$};
   \node (B) at (0.7,1.8) {$IIX$};
   \node (C) at (-1,2.8) {$XXX$};
   \node (D) at (1,2.8) {$YYX$};
   \node (E) at (2.2,2.8) {$YXY$};
   \node (F) at (4.3,2.8) {$XYY$};
   \node (G) at (1.6,4.5) {$YII$};
   \node (H) at (1.6,1.1) {$XII$};
   \node (I) at (2.5,1.8) {$IIY$};
   \node (J) at (3.2,0) {$IXI$};
   \draw[-]
   (A) edge (B)
   (B) edge (C)
   (C) edge [double] (D)
   (D) edge [double] (E)
   (D) edge (G)
   (E) edge (G)
   (E) edge [double] (F)
   (B) edge (D)
   (I) edge (E)
   (I) edge (J)
   (I) edge (H)
   (J) edge (H)
   (A) edge (H)
   (B) edge (H)
   (I) edge (F);
  \end{tikzpicture}

 \end{center}
\caption{A Mermin pentagram.}\label{penta}
\end{figure}
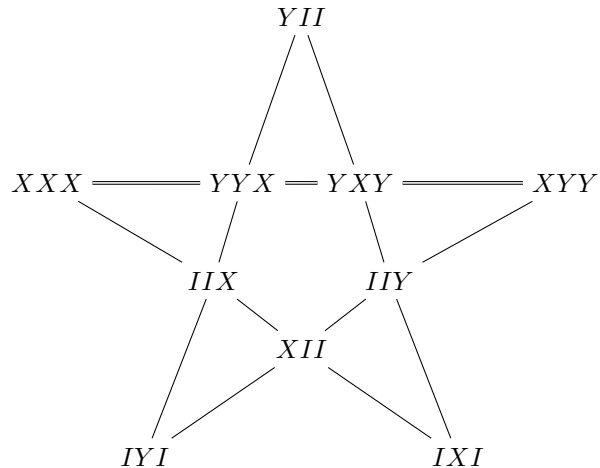

The Kochen-Specker (KS) theorem \cite{KS} is a fundamental result of quantum mechanics that rules out non-contextual hidden variables 
theories  by showing the impossibility to assign definite values to an observable independently of 
the context, i.\,e., independently of other compatible observables.  Many proofs have been proposed since the seminal
work of Kochen and Specker to simplify the initial argument based on the impossibility to color collections (bases) of rays in a $3$-dimensional space (see, e.\,g., \cite{WA1,WA2,Planat,Lison}). 
The observable-based KS-proofs  proposed by Peres \cite{Peres} and Mermin \cite{Mermin} in the 1990's 
provide a very simple and elegant version of KS-theorem, as we will now briefly recall. Let $X, Y, Z$ stand for the $2\times 2$ Pauli 
matrices and let $I$ be the identity matrix:
\begin{equation}
 X=\begin{pmatrix}
    0 & 1\\
    1& 0
   \end{pmatrix},~ Y=\begin{pmatrix}
   0 &-i\\
   i & 0 
   \end{pmatrix},~ Z=\begin{pmatrix}
   1 & 0\\
   0 & -1
   \end{pmatrix},~ I=\begin{pmatrix}
1 & 0\\
0 & 1
   \end{pmatrix}.
\end{equation}
Let us denote by $AB$ the tensor product $A\otimes B$ of two matrices from the above-given set. Then the  Mermin-Peres square depicted in Figure \ref{grid} provides an observable-based proof of the KS-theorem.
Each line of this $3 \times 3$ grid, i.\,e., each context, comprises mutually commuting 
operators and each operator squares to identity meaning that its eigenvalues are $\pm 1$. Next,  the product of 
the operators on each line is $II={\bf Id}$ except for one (shown in bold) 
where this product yields $-II=-{\bf Id}$. It is, however, clear that there is no way to assign a definite value $\pm 1$ to each operator to reproduce these product rules because each operator appears in exactly two lines/contexts. 
Another famous example of an operator-based KS-proof is furnished by the so-called Mermin pentagram, whose representative is shown in Figure \ref{penta}.
\noindent
 In this configuration the lines are made of four mutually commuting three-qubit operators. Again, each operator squares 
 to identity and the product of the operators on a given line is $\pm III=\pm {\bf Id}$. The odd number 
 of $-III=-{\bf Id}$ lines, in this example only one, leads to the 
 same contradiction as in the previous proof.
 
 Given these two examples, it is rather straightforward to see that an operator-based KS-proof relies on a configuration of operators satisfying the following properties:
 \begin{enumerate}
  \item The lines of the configuration consist of mutually commuting operators; such a line is called a context.
  \item All operators square to identity.
  \item All operators belong to an even number of contexts.
  \item The product of the operators on the same context is $\pm {\bf Id}$.
  \item There is an odd number of contexts giving $-{\bf Id}$.
 \end{enumerate}

\noindent
 In what follows we will focus on more specific contextual configurations, satisfying the following constrains:
 \begin{definition}\label{contextual}
 A configuration of operators is called a contextual $2$-configuration  if and only if
  \begin{enumerate}
  \item[1'.] The lines of the configuration consists of $p$ mutually commuting operators. 
  \item[2.] All operators square to identity.
  \item[3'.] All operators belong to exactly $2$ contexts.
  \item[4.] The product of the operators on the same context is $\pm {\bf Id}$. 
  \item[5.] There is an odd number of contexts giving $-{\bf Id}$.
 \end{enumerate}
 \end{definition}
 The definition of contextual $2$-configurations given by postulates 1', 2, 3', 4 and 5 is more restrictive than the
 one given by 1, 2, 3, 4 and 5. However, it is clear that the Mermin-Peres square and the Mermin 
 pentagram do satisfy those restrictive conditions.
Because  the product of observables on each context is $\pm {\bf Id}$, it follows 
that two contexts cannot have more than $p-2$ elements in common. 
\begin{definition}
A $2$-context-geometry will be a configuration of points/observables and lines/contexts such that:
\begin{enumerate}
 \item Each context contains the same number of points.
 \item Each point belongs to exactly two contexts.
\end{enumerate}
A configuration featuring $p$ points per context and $l$ contexts will be called an $(l,p)$-$2$-context-geometry.
\end{definition}

The Mermin-Peres square and the Mermin pentagram are, respectively, $(6,3)$- and $(5,4)$-$2$-context-geometries and, as we have seen, they are both contextual in the sense of Definition \ref{contextual}. 
It is, therefore, natural to ask if there exist smaller $2$-context-geometries furnishing an observable-based KS-proof. 

To address this question one first notes that the number of observables/points of an $(l,p)$-$2$-context-geometry is  $\dfrac{lp}{2}$. 
A Mermin-Peres square features $\dfrac{6\times 3}{2}=9$ observables, a Mermin Pentagram $\dfrac{5\times 4}{2}=10$ ones. Smaller $2$-context-geometries should thus be composed 
of $l\leq 5$ contexts, each context having $3\leq p\leq 4$ observables\footnote{The $p=2$ case can easily be ruled out because, up to a sign, only two operators $A$ and $A^{-1}$ will occur in such a configuration} such that $lp$ 
is even and $p\leq l$. 
Thus the only cases we need to consider are 
$(l,p)\in \{(5,4),(6,3),(4,4), (4,3)\}$. 
We shall proceed in two steps: first to enumerate all possible $2$-context-geometries and then to check whether such $2$-context-geometries are contextual. To check 
if a 
$2$-context-geometry is contextual, we label the points of the configuration by observables and  
simply compute the product of observables on each context. If the product of all the contexts gives $+{\bf Id}$, the corresponding $2$-context geometry is not  contextual.
Note that a similar argument is given in \cite{cleve}, referring to an original idea of F. Speelman. A more sophisticated 
version of this argument can be found in \cite{arkhipov}, where a graph-theoretical criteria is proposed to recognize 
a contextual configuration. However, for the cases considered in this note our approach is more efficient 
because it allows us to see why a particular configuration cannot be contextual or, when the configuration is potentially 
contextual, it also gives a hint of how to provide a realization of the configuration with multi-qubit observables. For example, we will see that both the Mermin-Peres square and the Mermin pentagram are, 
when embedded into symplectic polar spaces $W(3,2)$ and $W(5,2)$ underlying commutation relations between elements of the two-qubit respectively three-qubit Pauli group (see, e.\,g., \cite{san,plasan,havsan,thas}),  $2$-context-geometries that are always contextual  provided that just first four postulates of Definition \ref{contextual} are satisfied; in other words, for these two configurations constraint 5, {\it viz.} an odd number of contexts yielding ${\bf -Id}$, is the {\it consequence} of the remaining constrains.

In this note we do not consider configurations with contexts of varying size; these are discussed, for example, in \cite{WA3}. 
Our approach can be regarded as a combinatorial alternative to a group-theoretically-slanted program proposed recently by M. Planat 
\cite{Planat1,Planat2,Planat3, Planat4}, whose central objects are Grothendieck's {\it dessins d'enfants}.

The paper is organized as follows. In Sec. \ref{sec10} we enumerate and analyze all possible $(l,p)$-$2$-context-geometries 
with less than $10$ points and prove that only the $(6,3)$- and $(5,4)$-types  are suitable to furnish  contextual 
$2$-configurations, i.\,e., that the Mermin-Peres square and the Mermin pentagram are the {\it only} configurations 
with less than $10$ observables and the same number of observables per context that provide operator-based KS-proofs. 
In Sec. \ref{sec14} we describe a potentially contextual configuration featuring 
$12$ observables and give an example of a contextual configuration with $14$ observables. Finally, Sec. \ref{conclu}
is dedicated to concluding remarks.

\section{$2$-context-geometries having at most $10$ observables}\label{sec10} 
In this section we will analyze all possible $2$-context-geometries with less than or equal to $10$ points, i.\,e., consider 
$(l,p)$-$2$-context-geometries such that 
\begin{equation} (l,p)\in \{(4,3),(4,4),(6,3),(5,4)\}.\end{equation}
\subsection{The $(4,3)$-$2$-context-geometry, {\it aka} the Pasch configuration}
In the case of $p=3$, the associated $(4,3)$-$2$-context-geometry is a partial linear space, i.\,e. two contexts share 
at most one point/observable. The only configuration possessing $4$ lines, with $3$ points per line and $2$ lines per point, is 
the Pasch configuration \cite{enc}, well known in finite geometry for its role in classifications of Steiner triple systems (see. e.\,g., \cite{fgg}). It is easy to see that this configuration is unique by considering 
the configuration-matrix $\mathcal{M}_{(l,p)}$, an  $l\times l$ matrix where row $i$ and 
column $j$ represent, respectively, 
the context $C_i$ and the context $C_j$. The entry $m_{ij}$ is an integer that gives the number of operators shared by 
the contexts $C_i$ and $C_j$. By convention, we assume that $m_{ii}=0$. From  Definition \ref{contextual} it follows 
that configuration-matrices $\mathcal{M}_{(l,p)}$ are symmetric, 
$m_{ij} \in \{0,\dots,p-2\}$ and the sum of the entries in a given row and/or a column equals $p$. There is only one configuration-matrix $\mathcal{M}_{(4,3)}$,
\begin{equation}
 \mathcal{M}_{(4,3)}=\begin{pmatrix}
                      0 & 1 & 1 & 1\\
                      1 & 0 & 1 & 1\\
                      1 & 1 & 0 & 1\\
                      1 & 1 & 1 & 0
                     \end{pmatrix},
\end{equation}
which indeed corresponds to the Pasch configuration.

\begin{figure}[!h]
 \begin{center}
  \begin{tikzpicture}[scale=2]
   \node (A) at (0,0) {$A_1$};
\node (B) at (1,0) {$A_2$};
\node (C) at (2,0) {$A_3$};
\node (D) at (1,1.73) {$A_5$};
\node (E) at (0.5,0.866) {$A_6$};
\node (F) at (1.5,0.866) {$A_4$};
\draw[-]
(A) edge (B)
(B) edge (C)
(C) edge (F)
(F) edge (D)
(D) edge (E)
(E) edge (A)
(B) edge [bend left] (E)
(E) edge [bend left] (F)
(F) edge [bend left] (B);
  \end{tikzpicture}
\caption{The Pasch configuration.}\label{pash}
 \end{center}
\end{figure}
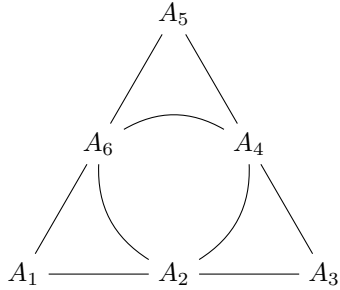

Let us assume that such a configuration, illustrated in Figure \ref{pash} in its most symmetric rendering,  is potentially contextual and label its six points by observables $A_1, A_2,\dots, A_6$.
\noindent
To find out whether this configuration is contextual, we calculate the product of observables along each line/context employing their associativity:
\begin{equation}
 (A_1A_2A_3)(A_3A_4A_5)(A_5A_6A_1)(A_2A_4A_6)=A_1A_2A_4A_6A_1A_2A_4A_6.
\end{equation}
Although the product of observables is, in general, not an observable, here the product of $A_1A_2A_4A_6$ is an observable. 
This is easy to see. As $A_1, A_2$ and $A_3$ are on the same context, $A_1A_2=\pm A_3$ and, similarly, $A_4A_6=\pm A_2$. But the same reasoning shows that $(\pm A_3)(\pm A_2)=\pm A_1$, i.\,e., 
$A_1A_2A_3A_4=\pm A_1$.
Therefore, we get 
\begin{equation}
\begin{array}{lll}
  (A_1A_2A_3)(A_3A_4A_5)(A_5A_6A_1)(A_2A_4A_6) & = & (A_1A_2A_4A_6)(A_1A_2A_4A_6)\\
                                               & = & (\pm A_1)^2\\
                                               &= & + {\bf Id},
                                               \end{array}
\end{equation}
meaning  that we cannot get an odd number of negative contexts; hence, the Pasch configuration is {\it not} contextual. 

\subsection{Two $(4,4)$-$2$-context-geometries}

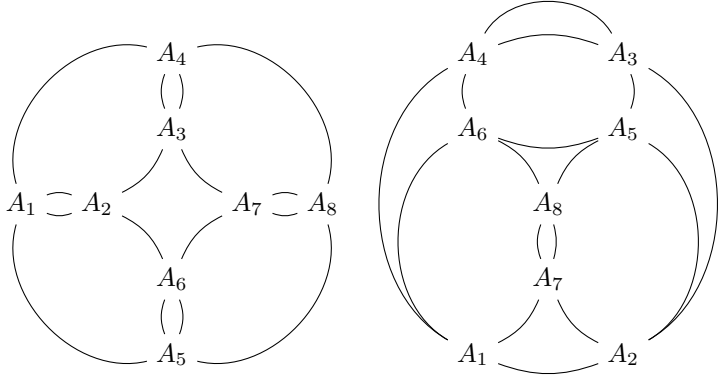
\begin{figure}[!h]
\begin{center}
\begin{tikzpicture}
 \node (A) at (0,0) {$A_1$};
 \node (B) at (1,0) {$A_2$};
 \node (C) at (2,1) {$A_3$};
 \node (D) at (2,2) {$A_4$};
 \node (E) at (3,0) {$A_7$};
 \node (F) at (4,0) {$A_8$};
 \node (G) at (2,-1) {$A_6$};
 \node (H) at (2,-2) {$A_5$};

 \node (AA) at (6,-2) {$A_1$};
 \node (BB) at (8,-2) {$A_2$};
 \node (CC) at (8,2) {$A_3$};
 \node (DD) at (6,2) {$A_4$};
 \node (EE) at (7,-1) {$A_7$};
 \node (FF) at (7,0) {$A_8$};
 \node (GG) at (6,1) {$A_6$};
 \node (HH) at (8,1) {$A_5$};
 \draw[-]
 (A) edge [bend right=20] (B)
 (B) edge [bend right=20] (C)
 (C) edge [bend right=20] (D)
 (D) edge [bend right=60] (A)
 (H) edge [bend left=20] (G)
 (G) edge [bend left=20] (E)
 (E) edge [bend left=20] (F)
 (F) edge [bend left=60] (H)
 (A) edge [bend left=20] (B)
 (B) edge [bend left=20] (G)
 (G) edge [bend left=20] (H)
 (H) edge [bend left=60] (A)
 (D) edge [bend right=20] (C)
 (C) edge [bend right=20] (E)
 (E) edge [bend right=20] (F)
 (F) edge [bend right=60] (D)
 
  (AA) edge [bend right=20] (BB)
 (BB) edge [bend right=60] (CC)
 (CC) edge [bend right=20] (DD)
 (DD) edge [bend right=60] (AA)
 (AA) edge [bend right=20] (EE)
 (EE) edge [bend right=15] (FF)
 (FF) edge [bend right=20] (GG)
 (GG) edge [bend right=60] (AA)
 (BB) edge [bend left=20] (EE)
 (EE) edge [bend left=20] (FF)
 (FF) edge [bend left=20] (HH)
 (HH) edge [bend left=60] (BB)
 (GG) edge [bend right=20] (HH)
 (HH) edge [bend right=20] (CC)
 (CC) edge [bend right=60] (DD)
 (DD) edge [bend right=20] (GG)
 ;
\end{tikzpicture}
\caption{The two $(4,4)$-$2$-context-geometries; the one on the left is contained in the so-called Miquel configuration.}\label{miquel}
\end{center}
\end{figure}

As a $(4,4)$-$2$-context-geometry is endowed with $8$ observables, it is impossible that all the entries of the corresponding configuration-matrix are equal to $1$ and, so, such a configuration is not a linear space. 
In particular, by enumerating all possibilities by a ``Sudoku-like'' argument, we find that, up to isomorhism, there are only two distinct configuration-matrices:
\begin{equation}
 \mathcal{M}_{(4,4)}=\begin{pmatrix}
                      0 & 2 & 2 & 0\\
                      2 & 0 & 0 & 2\\
                      2 & 0 & 0 & 2\\
                      0 & 2 & 2 & 0
                     \end{pmatrix} \text{ and } \mathcal{M}_{(4,4)}'=\begin{pmatrix}
                     0 & 2 & 1 & 1\\
                     2 & 0 & 1 & 1\\
                     1 & 1 & 0 & 2\\
                     1 & 1 & 2 & 0
                     \end{pmatrix}.
\end{equation}
The corresponding $2$-context-geometries (Figure \ref{miquel})  are configurations made of circles and neither of them is contextual.
To verify this claim, we follow  the same line of reasoning as in the case of the Pasch configuration. 
For the `Miquelian' configuration, we immediately get
\begin{equation}
 (A_1A_2A_3A_4)(A_4A_3A_7A_8)(A_8A_7A_6A_5)(A_5A_6A_2A_1)= +{\bf Id},
\end{equation}
while for the second configuration, we need three more steps to arrive at the same result:
\begin{equation}
\begin{array}{lll} 
(A_1A_2A_3A_4)(A_4A_3A_5A_6)(A_6A_1A_7A_8)(A_8A_7A_2A_5) & =& (A_1A_2A_5)(A_1A_2A_5)\\
 & =& (A_1(\pm A_7A_8))(A_1(\pm A_7A_8))\\
 & = & (\pm A_6)^2\\
 & =& + {\bf Id}.
\end{array}
\end{equation}

\subsection{Two $(6,3)$-$2$-context-geometries}
Interestingly, also in the $(6,3)$-case there are only two different configuration-matrices,
\begin{equation}
 \mathcal{M}_{(6,3)}=\begin{pmatrix}
                      0 & 1 & 1 & 1 & 0 & 0 \\
                      1 & 0 & 0 & 0 & 1 & 1\\
                      1 & 0 & 0 & 0 & 1 & 1\\
                      1 & 0 & 0 & 0 & 1 & 1\\
                      0 & 1 & 1 & 1 & 0 & 0 \\
                      0 & 1 & 1 & 1 & 0 & 0 
                     \end{pmatrix} \text{ and } \mathcal{M}_{(6,3)}'=\begin{pmatrix}
             0 & 1 & 1 & 1 & 0 & 0 \\
             1 & 0 & 0 & 0 & 1 & 1\\
             1 & 0 & 0 & 1 & 1 & 0\\
             1 & 0 & 1 & 0 & 0 & 1\\
             0 & 1 & 1 & 0 & 0 & 1\\
             0 & 1 & 0 & 1 & 1 & 0
\end{pmatrix},
\end{equation}
and, hence, only two non-isomorphic $(6,3)$-$2$-context-geometries. One of them -- illustrated in Figure \ref{prism} -- is called a prism, or a double-triangle, in the language of Steiner triple systems \cite{fgg}, and the other is nothing but a grid underlying our celebrated Mermin-Peres proof.

\begin{figure}[!h]
 \begin{center}
  \begin{tikzpicture}
   \node (A) at (0,0) {$A_1$};
 \node (B) at (1,1) {$A_2$};
 \node (C) at (2,2) {$A_3$};
 \node (D) at (3,1) {$A_4$};
 \node (E) at (4,0) {$A_5$};
 \node (F) at (3.5,-0.5) {$A_6$};
 \node (G) at (2,-2) {$A_7$};
 \node (H) at (0.5,-0.5) {$A_8$};
 \node (I) at (2,4) {$A_9$};
\draw[-]
(A) edge (B)
(B) edge (C)
(C) edge (D)
(D) edge (E)
(E) edge (F)
(F) edge (G)
(G) edge (H)
(H) edge (A)
(B) edge (I)
(D) edge (I)
(H) edge (B)
(F) edge (D);
  \end{tikzpicture}
\caption{A prism.}\label{prism}
 \end{center}
\end{figure}
Using the labeling of a prism shown in Figure \ref{prism}, we readily find that
\begin{equation}
\begin{array}{lll}
 & ~ & (A_1A_2A_3)(A_3A_4A_5)(A_5A_6A_7)(A_7A_8A_1)(A_8A_2A_9)(A_9A_4A_6) = \\
& ~ & (A_1A_2A_4A_6)(A_1A_2A_4A_6) = + {\bf Id}\\
  \end{array}
\end{equation}
since $A_1A_2A_4A_6=\pm {\bf Id}$; hence, a prism is not contextual.
As for a grid, the situation is more intricate. Employing its labeling depicted in Figure \ref{Mermin2}, we have
\begin{figure}[!h]
\begin{center}
\begin{tikzpicture}[scale=1.5]
\node (A) at (0,2) {$A_1$};
\node (B) at (1,2) {$A_2$};
\node (C) at (2,2) {$A_3$};
\node (D) at (0,1) {$A_4$};
\node (E) at (1,1) {$A_5$};
\node (F) at (2,1) {$A_6$};
\node (G) at (0,0) {$A_7$};
\node (H) at (1,0) {$A_8$};
\node (I) at (2,0) {$A_9$};
\draw[-]
(A) edge (B)
(A) edge (D)
(B) edge (C)
(B) edge (E)
(C) edge (F)
(D) edge (E)
(D) edge (G)
(E) edge (F)
(E) edge (H)
(F) edge (I)
(G) edge (H)
(H) edge (I);
\end{tikzpicture}
\end{center}
 \caption{A grid.}\label{Mermin2}
\end{figure}
\begin{equation}
\begin{array}{lll}
 & ~ &
 (A_1A_2A_3)(A_3A_6A_9)(A_9A_8A_7)(A_7A_4A_1)(A_4A_5A_6)(A_2A_5A_8)= \\
& ~ & (A_1A_2A_6A_8)(A_1A_6A_2A_8),
 \end{array}
\end{equation}
which implies that if  $A_2A_6=-A_6A_2$ then the grid-configuration is contextual. Therefore, if the product of the observables on 
each context is $\pm {\bf Id}$ and if the observables that are not on the same context anti-commute, 
then we are sure to 
have a contextual grid. But these two properties are naturally satisfied by observables associated with grids contained in the $N$-qubit Pauli groups, $N \geq 2$, when the latter are regarded as symplectic polar spaces $W(2N-1,2)$ of rank $N$ and order two  \cite{san}--\cite{thas}. In other words, a grid is always contextual when it is a subgeometry of a generalized Pauli group.

 \subsection{Four $(5,4)$-$2$-context-geometries}
A configuration-matrix analysis shows that apart from the Mermin pentagram there are other three $(5,4)$-$2$-context-geometries. 
However, unlike the pentagram, these are not partial linear spaces as their corresponding configuration-matrices feature entries from the set $\{0,1,2\}$; a representative of each of them is sketched in Figure \ref{pnt}.

\begin{figure}[!h]
 \begin{center}
  \begin{tikzpicture}
   \node (A) at (0,0) {$A_1$};
 \node (B) at (1,1) {$A_2$};
 \node (C) at (2,1) {$A_3$};
 \node (D) at (3,0) {$A_4$};
 \node (E) at (4,1) {$A_5$};
 \node (F) at (2.5,2) {$A_6$};
 \node (G) at (2,3) {$A_7$};
 \node (H) at (2,4) {$A_8$};
 \node (I) at (1,2) {$A_9$};
 \node (J) at (0,3) {$A_{10}$};
 
   \node (AA) at (8,0) {$A_1$};
 \node (BB) at (9,1) {$A_2$};
 \node (CC) at (10,1) {$A_3$};
 \node (DD) at (11,0) {$A_4$};
 \node (EE) at (11,2) {$A_5$};
 \node (FF) at (11.5,3) {$A_6$};
 \node (GG) at (12.5,3.5) {$A_7$};
 \node (HH) at (6.5,3.5) {$A_8$};
 \node (II) at (8,2) {$A_9$};
 \node (JJ) at (7.5,3) {$A_{10}$};
\draw[-]
(A) edge [bend right=20] (B)
 (B) edge [bend right=10] (I)
 (I) edge [bend right=20] (J)
 (J) edge [bend right=60] (A)
 (A) edge [bend left=20] (B)
 (B) edge [bend left=10] (C)
 (C) edge [bend left=20] (D)
 (D) edge [bend left=60] (A)
 (C) edge [bend left=20] (F)
 (F) edge [bend left=20] (E)
 (E) edge [bend left=20] (D)
 (D) edge [bend left=20] (C)
 (E) edge [bend left=20] (F)
 (F) edge [bend left=20] (G)
 (G) edge [bend left=20] (H)
 (H) edge [bend left=60] (E)
 (I) edge [bend right=20] (G)
 (G) edge [bend right=10] (H)
 (H) edge [bend right=50] (J)
 (J) edge [bend right=20] (I)
 
 (DD) edge [bend left=20] (EE)
  (EE) edge [bend left=10] (FF)
  (FF) edge [bend left=20] (GG)
  (GG) edge [bend left=60] (DD)
 (AA) edge [bend left=20] (BB)
  (BB) edge [bend left=10] (CC)
  (CC) edge [bend left=20] (DD)
  (DD) edge [bend left=60] (AA)
 (AA) edge [bend right=10] (II)
 (II) edge [bend right=10] (JJ)
 (JJ) edge [bend right=20] (HH)
 (HH) edge [bend right=60] (AA)
 (HH) edge [bend right=10] (JJ)
  (JJ) edge [bend right=10] (FF)
  (FF) edge [bend right=10] (GG)
  (GG) edge [bend right=60] (HH)
  (BB) edge [bend left=20] (II)
  (II) edge [bend left=30] (EE)
  (EE) edge [bend left=20] (CC)
  (CC) edge [bend left=20] (BB)
;
  \end{tikzpicture}
 \end{center}
\end{figure}
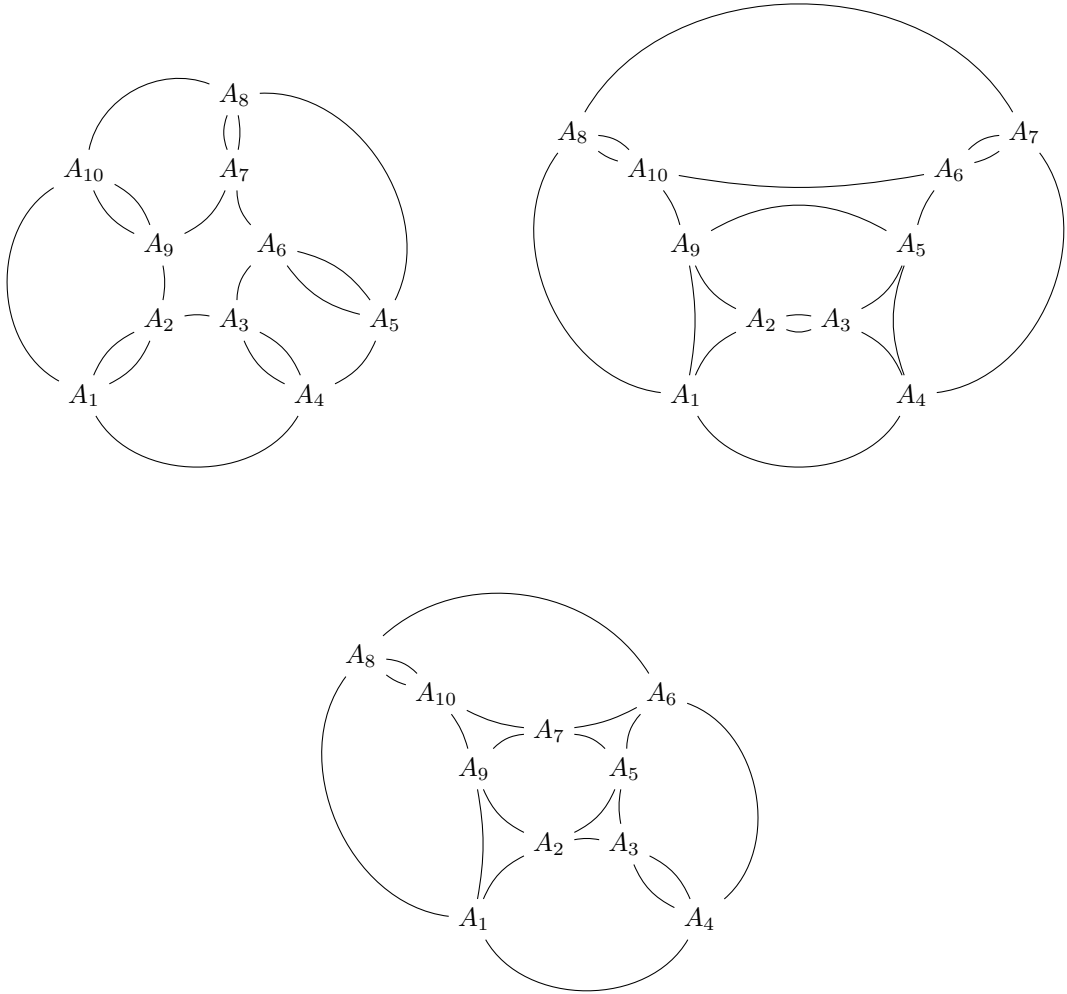
\begin{figure}[!h]
 \begin{center}
  \begin{tikzpicture}
   \node (A) at (0,0) {$A_1$};
 \node (B) at (1,1) {$A_2$};
 \node (C) at (2,1) {$A_3$};
 \node (D) at (3,0) {$A_4$};
 \node (E) at (2,2) {$A_5$};
 \node (F) at (2.5,3) {$A_6$};
 \node (G) at (1,2.5) {$A_7$};
 \node (H) at (-1.5,3.5) {$A_8$};
 \node (I) at (0,2) {$A_9$};
 \node (J) at (-0.5,3) {$A_{10}$};
\draw[-]
(D) edge [bend left=20] (C)
   (C) edge [bend left=10] (E)
  (E) edge [bend left=20] (F)
  (F) edge [bend left=60] (D)
  (A) edge [bend left=20] (B)
   (B) edge [bend left=10] (C)
   (C) edge [bend left=20] (D)
   (D) edge [bend left=60] (A)
  (A) edge [bend right=10] (I)
  (I) edge [bend right=10] (J)
  (J) edge [bend right=20] (H)
  (H) edge [bend right=60] (A)
 (B) edge [bend right=20] (E)
  (E) edge [bend right=20] (G)
  (G) edge [bend right=20] (I)
   (I) edge [bend right=20] (B)
  (H) edge [bend right=10] (J)
   (J) edge [bend right=10] (G)
   (G) edge [bend right=10] (F)
   (F) edge [bend right=50] (H)
;
  \end{tikzpicture}
\caption{Three nonisomorphic $(5,4)$-$2$-context geometries that are not partial linear spaces.}\label{pnt}
 \end{center}
\end{figure}
We leave it as an exercise for the interested reader to verify that none of these three configurations is contextual. It is also worth mentioning that the pentagram is, like the grid, always 
contextual when being a subgeometry of a multi-qubit symplectic polar space, as
the sole requirement that two observables commute/anti-commute if they are/are not collinear guarantees that there are an odd number of contexts whose product is $-{\bf Id}$. 

\section{Some 2-context-geometries with $12$ and $14$ observables}\label{sec14}
At this point it is natural to ask: What is the next $2$-context-geometry in the hierarchy that provides an observable-based KS-proof? As it is obvious that there is no such geometry with $11$ observables, one has to search for it among $12$-point configurations.
\subsection{A potentially contextual 2-context-geometry with $12$ observables}
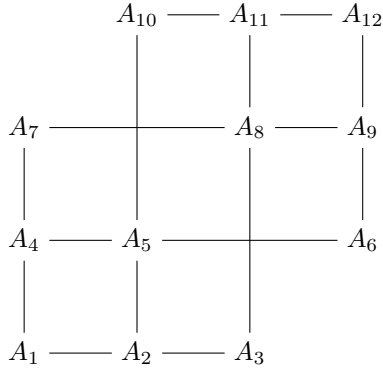
\begin{figure}[!h]
\begin{center}
\begin{tikzpicture}[scale=1.5]
\node (A) at (0,0) {$A_1$};
\node (B) at (1,0) {$A_2$};
\node (C) at (2,0) {$A_3$};
\node (D) at (0,1) {$A_4$};
\node (E) at (1,1) {$A_5$};
\node (F) at (3,1) {$A_6$};
\node (G) at (0,2) {$A_7$};
\node (H) at (2,2) {$A_8$};
\node (I) at (3,2) {$A_9$};
\node (J) at (1,3) {$A_{10}$};
\node (K) at (2,3) {$A_{11}$};
\node (L) at (3,3) {$A_{12}$};
\draw[-]
(A) edge (B)
(B) edge (C)
(A) edge (D)
(B) edge (E)
(C) edge (H)
(D) edge (E)
(D) edge (G)
(E) edge (F)
(E) edge (J)
(F) edge (I)
(G) edge (H)
(H) edge (I)
(I) edge (L)
(L) edge (K)
(K) edge (J)
(H) edge (K);
\end{tikzpicture}
\end{center}
 \caption{A highly-symmetric $(8,4)$-$2$-context-geometry that is (potentially) contextual.}\label{compov}
\end{figure}

There are, indeed, several $(8,4)$-$2$-context-geometries which are `potentially' contextual. The adjective `potentially' here means that such geometry satisfies all the constrains of Definition \ref{contextual}, but 
we have been so far unable to find its explicit labeling in terms of elements of some multi-qubit Pauli group.
An illustrative example of such geometry is provided by the complement of an ovoid of a $4 \times 4$-grid, portrayed in Figure \ref{compov}.

If we again assume that non-colinear observables anti-commute, then from the labeling of Figure \ref{compov} we readily ascertain that:
\begin{equation}
\begin{array}{l}
 (A_1A_2A_3)(A_3A_8A_{11})(A_{10}A_{11}A_{12})(A_{12}A_9A_6)(A_6A_5A_4)(A_1A_4A_7)(A_7A_8A_9)(A_2A_5A_{10})\\
 =(A_1A_2A_8A_{10}A_9A_5)(A_1A_8A_9A_2A_5A_{10})=-(A_1A_2A_8A_{10}A_9A_5)^2=-{\bf Id},
 \end{array}
\end{equation}
the last equality stemming from the fact that the product $A_1A_2A_8A_{10}A_9A_5$ is an observable. It is a challenging question to see whether there indeed exists a realization of this configuration in terms of elements of a certain multi-qubit Pauli group.


\subsection{A contextual $2$-context-geometry with $14$ observables}
As there is no $2$-context-geometry with $13$ points, the next case in the hierarchy are geometries endowed with $14$ observables.
One of the most prominent of them, which is contextual and for which we succeeded in finding an explicit realization in terms of the four-qubit Pauli group, is a self-intersecting heptagon of Schl\"afli symbol $\{7/2\}$, depicted in Figure \ref{hepta}.\footnote{It is worth mentioning that this heptagram can also be found in a noteworthy $21_4$-configuration discovered by Felix Klein as early as 1879 and studied in detail in the real plane by Gr\"unbaum and Rigby \cite{gr}.} 
\begin{figure}[!h]
\begin{center}
\begin{tikzpicture}[scale=2]
\node (A) at (0,0) {$IYII$};
\node (B) at (0.46,0.9) {$IIXI$};
\node (C) at (0,1.78) {$YXXI$};
\node (D) at (1,2) {$XYXY$};
\node (E) at (1.38,2.88) {$XIIY$};
\node (F) at (2.16,2.26) {$XXZI$};
\node (G) at (3.1,2.46) {$YYZY$};
\node (H) at (3.1,1.5) {$IIZY$};
\node (I) at (3.88,0.9) {$IXII$};
\node (J) at (3.1,0.28) {$IIYZ$};
\node (K) at (3.1,-0.7) {$YYYZ$};
\node (L) at (2.14,-0.48) {$IIYI$};
\node (M) at (1.38,-1.1) {$IXIZ$};
\node (N) at (0.96,-0.2) {$YIIZ$};
\draw[-]
(A) edge (B)
(B) edge (C)
(B) edge (D)
(C) edge [double] (D)
(D) edge [double] (F)
(D) edge (E)
(E) edge (F)
(F) edge [double] (G)
(G) edge (H)
(F) edge (H)
(H) edge (I)
(I) edge (J)
(H) edge (J)
(J) edge (K)
(K) edge (L)
(J) edge (L)
(L) edge (M)
(M) edge (N)
(L) edge (N)
(N) edge (A)
(N) edge (B);
\end{tikzpicture}
\end{center}
 \caption{A `magic' heptagram of four-qubit observables.}\label{hepta}
\end{figure}

Our heptagram belongs to a large family of regular star polygons. A $\{p/q\}$ regular star polygon, with $p,q$ being positive integers,
is obtained from a $p$-regular polygon by joining every $q$th vertex of the polygon. The pentagram is the first regular star polygon of 
 Schl\"afli symbol $\{5/2\}$. Regular star polygons are self-intersecting and, if also all points of self-intersections are included, they form a remarkable sequence of $2$-context-geometries with $pq$ points.  It would, therefore, be desirable to clarify which of them are potentially contextual and, as a next step, to address the question of realizability of the latter in terms of the symplectic geometry of multi-qubit Pauli groups.

\section{Conclusion}\label{conclu}
We have outlined a rather elementary algebraic-geometrical recipe for ascertaining which point-line configurations can serve as observable-based 
proofs of the Kochen-Specker Theorem. It was proved that under the assumption that every context contains the same number of observables and 
that every observable belongs to exactly two contexts, the simplest such configurations, in terms of the number of points/observables,
are the celebrated Mermin-Peres square and Mermin pentagram. 
We also pointed out that when these configurations are viewed as substructures of symplectic polar spaces underlying multi-qubit Pauli groups, they are automatically 
contextual, in the sense that constraint 5 of Definition \ref{contextual} is always satisfied.
The next contextual configuration was found to possess 12 observables, though we have not yet been able to find its explicit realization in terms of elements of a certain multi-qubit Pauli group.  This was, however, possible for a 14-point $\{7/2\}$-heptagram in terms of four-qubit observables. On the other hand, we have also demonstrated why some other prominent finite geometries, like the Pasch configuration and the prism playing a crucial role in classifying Steiner triple systems, are not contextual.
Last but not least, there is an important byproduct of our reasoning, namely the necessity to deepen our understanding of the fine structure of symplectic polar spaces of multiple-qubit Pauli groups in order to be able to tackle more efficiently the question of explicit realizations of contextual configurations.


\section*{Acknowledgments}
This work was supported by  the French Conseil R\'egional Research Project RECH-MOB15-000007 and, in part, by the Slovak VEGA Grant Agency, Project No. 2/0003/16.
 

\end{document}